\documentclass[12pt]{article}
\usepackage{amsfonts}
\usepackage{eurosym}
\usepackage{graphicx}
\usepackage[utf8]{inputenc}
\usepackage[T1]{fontenc}
\usepackage{indentfirst}
\usepackage[margin=0.6in,nomarginpar]{geometry}
\usepackage[final]{hyperref}
\usepackage{amsmath}
\usepackage{hyperref}
\usepackage{cite}
\usepackage{subcaption}
\usepackage{caption}
\usepackage{amssymb}
\usepackage{authblk}
\usepackage{multirow}
\usepackage[table]{xcolor}
\usepackage{orcidlink}

\hypersetup{
colorlinks=true,
linkcolor=blue,
citecolor=blue,
filecolor=magenta,
urlcolor=blue
}

\begin{document}

\title{A Path Integral Treatment of  Time-dependent Dunkl Quantum Mechanics}
\author{A. Benchikha\orcidlink{0009-0003-9848-0822} \thanks{
benchikha4@yahoo.fr/amar.benchikha@umc.edu.dz} \\
$^{1}$D\'{e}partement de EC, Facult\'{e} de SNV, Universit\'{e} Constantine
1 Fr\`{e}res Mentouri,\\
Constantine, Algeria.\\
$^{2}$Laboratoire de Physique Math\'{e}matique et Subatomique,\\
LPMS, Universit\'{e} Constantine 1 Fr\`{e}res Mentouri, Constantine,
Algeria. \and B. Hamil\orcidlink{0000-0002-7043-6104} \thanks{%
hamilbilel@gmail.com/bilel.hamil@umc.edu.dz} \\
Laboratoire de Physique Math\'{e}matique et Subatomique,\\
LPMS, Universit\'{e} Constantine 1 Fr\`{e}res Mentouri, Constantine,
Algeria. \and B. C. L\"{u}tf\"{u}o\u{g}lu\orcidlink{0000-0001-6467-5005} 
\thanks{%
bekir.lutfuoglu@uhk.cz (Corresponding author)}, \\
Department of Physics, Faculty of Science, University of Hradec Kralove,\\
Rokitanskeho 62/26, Hradec Kralove, 500 03, Czech Republic.}
\date{\today }
\maketitle

\begin{abstract}
This study addresses a significant gap in the literature by extending the path integral formalism to time-dependent systems within the Wigner-Dunkl framework, deriving exact propagator solutions for analytically solvable cases. By employing generalized canonical transformations, we reformulated the path integral to develop an explicit expression for the propagator. This formalism is applied to specific cases, including a Dunkl-harmonic oscillator with time-dependent mass and frequency. Solutions for the Dunkl-Caldirola-Kanai oscillator and a model with a strongly pulsating mass are derived, providing exact propagator expressions and corresponding wave functions. These findings extend the utility of Dunkl operators in quantum mechanics, offering new insights into the dynamics of time-dependent quantum systems and possibly find application in quantum optics, plasma physics, and other fields.
\end{abstract}

\section{Introduction}

Recently, there has been a significant interest in the exploration of time-dependent quantum systems \cite{Onah, Guasti, Herrera, Benchikha, Dutta, Pinaki, Ashot, Ganguly, Tibaduiza, Badri}. This is because these systems have important applications in various fields of physics, including quantum optics \cite{Berger}, cosmology \cite{Colegrave}, nanotechnologies \cite{Brown}, and plasma physics \cite{Bohn}. The diverse applications of these systems have prompted the development of various methods and techniques to study such systems. Notable among these are the Lewis-Riesenfeld invariant operator method \cite{Lewis, Riesenfeld}, path integral formalism \cite{Chetouani1}, and coherent states \cite{Zhang}. The Lewis-Riesenfeld invariant operator method is an approach that enables the determination of a complete set of solutions to the Schr\"{o}dinger equation for time-dependent potentials by utilizing the eigenstates of an invariant operator. This method consists of two main steps: first, constructing the invariant operator, and second, expressing Schr\"{o}dinger's wave function using the eigenstates of this invariant operator, combined with a time-dependent phase factor. On the other hand, the Feynman path integral formalism provides an approach to solving quantum mechanical problems \cite{Feynman} alternative to the well-known methods of Heisenberg and Schr\"{o}dinger. However, this formulation's applicability has been limited because explicit expressions for propagators are available only for a few cases. In the path integral approach, various time-dependent quantum systems have been studied, such as the harmonic oscillator with varying frequency or mass \cite{Badri, Chetouani1, Khandekar, Gerry, Dhara, Cheng, Winotai, Sabir}, the infinite potential well \cite{Chetouani1}, and the time-dependent model of the form $V\left( x-f\left( t\right) \right) $ \cite{Duru}.

In 1950, Wigner studied the relationship between quantum dynamics and commutation relations. He demonstrated that the idea of quantization, represented by the canonical commutation relation $\left[ x,p\right] =i,$ is a solution to the Heisenberg equation of motion, but it is not the only possible solution \cite{Wigner}. Just a year later, Yang considered the problem that had been addressed by Wigner \cite{Yang}. By using a more precise definition of Hilbert space and introducing another condition involving a rigorous theorem about expanding mathematical objects, Yang showed that commutation relations could be deduced uniquely if a reflection operator, $\hat{R}$, is taken into consideration:
\begin{equation}
p=\frac{1}{i}\left( \frac{d}{dx}+\frac{\nu }{x}\hat{R}\right) ,
\end{equation}
where $\nu $ denotes the Wigner parameter. In 1989, Dunkl introduced a new type of differential operator, \textbf{the Dunkl operator}, $D$, defined as a linear combination of partial derivatives weighted by the reflection operator, allowing the study of special functions and their properties in a more general setting \cite{Dunkl, Dunkl1}. The one-dimensional version of the Dunkl operator is given by
\begin{equation}
D=\frac{1}{i}\left( \frac{d}{dx}+\frac{\nu }{x}\left( 1-\hat{R}\right) \right) .
\end{equation}
Quantum mechanical systems involving the reflection operator have recently found applications in physics related, in particular, to anyons \cite{Mikhail, Peter}, supersymmetry \cite{Francisco, SPlyushchay, Correa, Nieto, Nieto1, Vit, Vit1}, integrable systems \cite{Francisco, Polychronakos}, non-commutative geometry \cite{Peter}, and PT-symmetry \cite{Luis, Alcala}. Substituting the ordinary partial derivative with the Dunkl derivative has found applications in a variety of physics fields, enabling the exact study of different quantum mechanics problems using various methods \cite{Vincent1, Vincent2, Vincent3, Mota1, Hassanabadi1, Hassanabadi2, Merad, Mota2, Hamil1, Hamil2, Hamil3, Mota3, Hassanabadi3, Hassanabadi4, Quesne1, Quesne2, Quesne3, Bouguerne, Hocine1, Hocine2, Schulze}. Recently, four papers have explored the development of path integral formalism within the context of Wigner-Dunkl quantum mechanics \cite{Benchikha2, Junker, Benzair1, Benzair2}. One of these papers focused on constructing the propagator for one-dimensional quantum systems incorporating the Dunkl derivative \cite{Benchikha2}. The authors employed the configuration space representation and Feynman's path integral approach in their analysis. Their findings revealed that the Dunkl path integral kernel can be expressed as a combination of even and odd kernel propagators.

Our aim in this paper is to explore the Dunkl path integral formalism for the propagator in the presence of a time-dependent potential. Unlike other approaches, such as the Lewis-Riesenfeld invariant method, the Dunkl formalism naturally integrates reflection symmetries, making it particularly suitable for systems with time-dependent potential and mass. The treatment is mainly based on suitably chosen generalized canonical transformations and auxiliary equations. This paper is organized as follows: In Sec. \ref{sec2}, we reformulate the path integral formalism for the propagator of the time-dependent system. In Sec. \ref{sec3}, we consider the case of the Dunkl-harmonic oscillator with a time-dependent mass and frequency, and we calculate the relative propagator and the wave functions. In Sec. \ref{sec4}, we illustrate the general procedure by considering two models of varying mass. Finally, Sec. \ref{concl} contains the conclusion.

\section{Time-dependent propagator}

\label{sec2}

We start by outlining a method to compute the path integral for
non-relativistic Wigner-Dunkl quantum systems with a time-dependent mass in the presence of general time-dependent potentials. The dynamics of the system are governed by the Wigner-Dunkl Hamiltonian, which incorporates time-dependent mass and potential terms as follows: 

\begin{equation}
H=-\frac{\hbar ^{2}}{2m\left( t\right) }D^{2}+V\left( x,t\right) ,
\label{phase}
\end{equation}
where 
\begin{equation}
D^{2}=\frac{d^{2}}{dx^{2}}+\frac{2\nu }{x}\frac{d}{dx}-\frac{\nu }{x^{2}}%
\left( 1-\hat{R}\right) .
\end{equation}
In the functional integral formalism, the Wigner-Dunkl-propagator
corresponding to the system described by Eq. (\ref{phase}) can be expressed
as \cite{Benchikha2,Junker}, 
\begin{equation}
K\left( x_{f};t_{f},x_{i};t_{i}\right)
=K_{+}\left(x_{f};t_{f},x_{i};t_{i}\right) +sng\left( x_{f}x_{i}\right)
K_{-}\left(x_{f};t_{f},x_{i};t_{i}\right) ,  \label{decom}
\end{equation}%
where $sng\left( x\right) $ denotes the sign function, $K_{+}$ represents
the even kernel 
\begin{equation}
K_{+}=\frac{1}{\left( y_{a}y_{b}\right) ^{\nu }}\int Dy\exp \left\{ \frac{i}{%
\hbar }\int_{0}^{T}dt\left[ \frac{m\left( t\right) }{2}\dot{y}^{2}-\frac{%
\hbar ^{2}\left( \alpha ^{2}-\frac{1}{4}\right) }{2m\left( t\right) y^{2}}%
-V(y,t)\right] \right\} ,\text{ with }\alpha =\nu -1/2,  \label{k+}
\end{equation}%
and $K_{-}$ corresponds to the odd kernel 
\begin{equation}
K_{-}=\frac{1}{\left( y_{a}y_{b}\right) ^{\nu }}\int Dy\exp \left\{ \frac{i}{%
\hbar }\int_{0}^{T}dt\left[ \frac{m\left( t\right) }{2}\dot{y}^{2}-\frac{%
\hbar ^{2}\left( \beta ^{2}-\frac{1}{4}\right) }{2m\left( t\right) y^{2}}%
-V(y,t)\right] \right\} ,\text{ with }\beta =\nu +1/2,  \label{k-}
\end{equation}
propagators. To keep the mass term in the kinetic energy constant, we can
convert the propagator into an appropriate form by continuing the treatment
in phase space as described in \cite{Chetouani1}. In this context, the
kernel propagator in phase space takes the form: 
\begin{equation}
K_{e}=\frac{1}{\left( y_{f}y_{i}\right) ^{\nu }}\int \mathcal{D}y\mathcal{DP}%
\exp \left[ \frac{i}{\hbar }\left( \mathcal{P}\dot{y}-\mathcal{H}\right) %
\right] ,  \label{2}
\end{equation}%
where 
\begin{equation}
\mathcal{H=}\frac{\mathcal{P}^{2}}{2m(t)}+\frac{\hbar ^{2}\left( \lambda
^{2}-\frac{1}{4}\right) }{2m\left( t\right) y^{2}}+V(y,t),
\end{equation}
and 
\begin{equation}
y=\left\vert x\right\vert , \quad \text{and} \quad \lambda =\nu -\frac{e}{2}.
\end{equation}
Here, $e$ takes the value $+1$ for the even case and $-1$ for the odd case.

Our purpose here is to transform the time-dependent mass and potential terms into a stationary form using a set of transformations. We employ the method of canonical transformations \cite{Chetouani1}, specifically performing a time-dependent canonical transformation $\left(y,\mathcal{P},t\right) \xrightarrow{F\left( Q, P; t\right)}\left( Q,P,t\right) $ defined as,

\begin{eqnarray}
y &=&Q\rho \left( t\right) ,  \label{trans1} \\
\mathcal{P} &=&\frac{P}{\rho \left( t\right) },  \label{trans2}
\end{eqnarray}%
with the generating function 
\begin{equation}
F(y,P;t)=P\frac{y}{\rho }=PQ,  \label{gene funct}
\end{equation}%
which can be derived from the following equations of classical mechanics%
\begin{equation}
\mathcal{P}=\frac{\partial }{\partial y}F(y,P;t),\qquad Q=\frac{\partial }{%
\partial P}F(y,P;t).  \label{cl eqs}
\end{equation}%
As a consequence of these transformations, we reformulate the action defined
in Eq. (\ref{2}) in terms of the new conjugate variables as follows: 
\begin{eqnarray}
\mathcal{P}\dot{y}-\mathcal{H} &=&P\dot{Q}-\mathit{H}+\frac{\partial }{%
\partial t}F^{\prime },\text{ }  \notag \\
&=&P\dot{Q}-\left( \frac{P^{2}}{2m\rho ^{2}}-\frac{PQ\dot{\rho}}{\rho }+%
\frac{\hslash ^{2}\left( \lambda ^{2}-\frac{1}{4}\right) }{2m\rho ^{2}Q^{2}}%
+V\left( \rho \,Q,t\right) \right) ,  \label{cl3}
\end{eqnarray}%
where%
\begin{equation}
\mathit{H}\left( Q,P,t\right) =\frac{P^{2}}{2m\rho ^{2}}-\frac{PQ\dot{\rho}}{%
\rho }+\frac{\hslash ^{2}\left( \lambda ^{2}-\frac{1}{4}\right) }{2m\rho
^{2}Q^{2}}+V\left( \rho \,Q,t\right) ,
\end{equation}%
is the new Hamiltonian which governs the movement in the system $\left(
Q,P,t\right) $, and%
\begin{equation}
F=F^{^{\prime }}+PQ.
\end{equation}
Here, the dot notation over the variable refers to the time derivative. The
transformation subsequently alters the integral measure; following the
procedures described in \cite{Chetouani1}, we find that the path integral
measure transforms as follows: 
\begin{equation}
\mathcal{D}y\mathcal{DP}=\frac{1}{\left( \rho _{f}\rho _{i}\right) ^{\frac{1%
}{2}}}\mathcal{D}Q\mathcal{D}P.  \label{mesure1}
\end{equation}%
From Eqs. (\ref{cl3}) and (\ref{mesure1}), it follows that the evolution of
the physical system in the new coordinate system is characterized by 
\begin{eqnarray}
K_{e} &=&\frac{1}{\left( \rho _{f}\rho _{i}\right) ^{\nu +\frac{1}{2}}\left(
Q_{f}Q_{i}\right) ^{\nu }}\int \mathcal{D}Q\mathcal{D}P\exp \Bigg\{\frac{i}{%
\hslash }\int\limits_{t_{i}}^{t_{f}}\Bigg[P\dot{Q}  \notag \\
&&\left. -\left[ \frac{P^{2}}{2m\left( t\right) \rho ^{2}}-\frac{PQ\dot{\rho}%
}{\rho }+\frac{\hslash ^{2}\left( \lambda ^{2}-\frac{1}{4}\right) }{2m\rho
^{2}Q^{2}}+V\left( \rho \,Q,t\right) \right] \right] dt\Bigg\}.
\label{propagator2}
\end{eqnarray}%
We now utilize a transformation of the time variable, changing $t$ to $s$: 
\begin{equation}
ds=\frac{dt}{\rho ^{2}\left( t\right) m\left( t\right) },
\label{time transf}
\end{equation}%
so that $K_{e}$ becomes 
\begin{eqnarray}
K_{e} &=&\frac{1}{\left( \rho _{f}\rho _{i}\right) ^{\nu +\frac{1}{2}}\left(
Q_{f}Q_{i}\right) ^{\nu }}\int \mathcal{D}Q\mathcal{D}P\exp \left\{ \frac{i}{%
\hslash }\int\limits_{s_{i}}^{s_{f}}\Bigg[P\dot{Q}\right.   \notag \\
&&\left. \left. -\left[ \frac{P^{2}}{2}-\frac{PQ\overset{\cdot }{\bar{\rho}}%
}{\bar{\rho}}+\frac{\hslash ^{2}\left( \lambda ^{2}-\frac{1}{4}\right) }{%
2Q^{2}}+\bar{m}\bar{\rho}^{2}V\left( \bar{\rho}\,Q,\int\limits^{s}\bar{m}%
\bar{\rho}^{2}d\sigma \right) \right] \right] ds\right\} ,
\label{propagator3}
\end{eqnarray}%
where $\overset{\cdot }{\bar{\rho}}=\frac{d\rho }{ds},$ \ $\bar{\rho}=\rho
\left( s\right) $,\ $\bar{m}$\ $=m\left( s\right) $. At this stage, we
notice the presence of an additional term $\frac{PQ\overset{\cdot }{\bar{\rho%
}}}{\bar{\rho}}$ in the propagator expression given in Eq. (\ref{propagator3}%
). To eliminate this term, we introduce the following canonical
transformation $\left( Q,P,s\right) \xrightarrow{F_{1}\left( Q, \mathbf{P};
s\right)}\left( \tilde{Q},\mathbf{P},s\right) $: 
\begin{equation}
\mathbf{P}=P-\tilde{Q}\frac{\overset{\cdot }{\bar{\rho}}}{\bar{\rho}},\text{
\ \ \ \ }Q=\tilde{Q},  \label{new transformation}
\end{equation}%
with the generating function given by; 
\begin{equation}
F_{1}\left( Q,\mathbf{P},s\right) =\mathbf{P}Q+Q^{2}\frac{\overset{\cdot }{%
\bar{\rho}}}{2\bar{\rho}}.  \label{new gf}
\end{equation}%
In this case, the motion of the particle within the $\left( Q,\mathbf{P}%
,s\right) $ system is governed by the new Hamiltonian: 
\begin{equation}
\mathbf{H=}\frac{\mathbf{P}^{2}}{2}+\frac{1}{2}\Omega ^{2}Q^{2}+\frac{%
\hslash ^{2}\left( \lambda ^{2}-\frac{1}{4}\right) }{2Q^{2}}+\bar{m}\bar{\rho%
}^{2}V\left( \bar{\rho}\,Q,\int\limits^{s}\bar{m}\bar{\rho}^{2}d\sigma
\right) ,  \label{newH}
\end{equation}%
where 
\begin{equation}
\Omega ^{2}=\frac{\overset{\cdot \cdot }{\bar{\rho}}}{\rho }-2\left( \frac{%
\overset{\cdot }{\bar{\rho}}}{\rho }\right) ^{2}=m^{2}\rho ^{3}\left( \ddot{%
\rho}+\frac{\dot{m}}{m}\dot{\rho}\right) .  \label{newp}
\end{equation}%
Thus, the propagator corresponding to the system in Eq. (\ref{newH}) can be
expressed as: 
\begin{eqnarray}
K_{e} &=&\frac{\exp \left[ \frac{i}{2\hslash }\left( \frac{\overset{\cdot }{%
\bar{\rho}_{f}}}{\rho _{f}}Q_{f}^{2}-\frac{\overset{\cdot }{\bar{\rho}_{i}}}{%
\rho _{i}}Q_{i}^{2}\right) \right] }{\left( \rho _{f}\rho _{i}\right) ^{\nu +%
\frac{1}{2}}\left( Q_{f}Q_{i}\right) ^{\nu }}\int \mathcal{D}Q\exp \left\{ 
\frac{i}{\hslash }\int \left[ \frac{\dot{Q}^{2}}{2}\right. \right.   \notag
\\
&&\left. \left. -\left[ \frac{\Omega ^{2}}{2}Q^{2}+\frac{\hslash ^{2}\left(
\lambda ^{2}-\frac{1}{4}\right) }{2Q^{2}}+\bar{m}\bar{\rho}^{2}V\left( \bar{%
\rho}\,Q,\int\limits^{s}\bar{m}\bar{\rho}^{2}d\sigma \right) \right] \right]
ds\right\} .  \label{propagatorfinal}
\end{eqnarray}%
We now decompose the kernel propagator into two parity components: the \textbf{even kernel} (\( K_+ \)) and the \textbf{odd kernel} (\( K_- \)),  distinguished by their respective eigenvalues, $e$. The propagator for
the even parity case is given by: 
\begin{eqnarray}
K_{+} &=&\frac{\exp \left[ \frac{i}{2\hslash }\left( \frac{\overset{\cdot }{%
\bar{\rho}_{f}}}{\rho _{f}}Q_{f}^{2}-\frac{\overset{\cdot }{\bar{\rho}_{i}}}{%
\rho _{i}}Q_{i}^{2}\right) \right] }{\left( \rho _{f}\rho _{i}\right) ^{\nu +%
\frac{1}{2}}\left( Q_{f}Q_{i}\right) ^{\nu }}\int \mathcal{D}Q\exp \left\{ 
\frac{i}{\hslash }\int\limits_{s_{i}}^{s_{f}}\left[ \frac{\dot{Q}^{2}}{2}%
\right. \right.   \notag \\
&&\left. \left. -\left[ \frac{\Omega ^{2}}{2}Q^{2}+\frac{\hslash ^{2}\left(
\alpha ^{2}-\frac{1}{4}\right) }{2Q^{2}}+\bar{m}\bar{\rho}^{2}V\left( \bar{%
\rho}\,Q,\int\limits^{s}\bar{m}\bar{\rho}^{2}d\sigma \right) \right] \right]
ds\right\} ,  \label{propagator+}
\end{eqnarray}%
and the odd parity case propagator is: 
\begin{eqnarray}
K_{-} &=&\frac{\exp \left[ \frac{i}{2\hslash }\left( \frac{\overset{\cdot }{%
\bar{\rho}_{f}}}{\rho _{f}}Q_{f}^{2}-\frac{\overset{\cdot }{\bar{\rho}_{i}}}{%
\rho _{i}}Q_{i}^{2}\right) \right] }{\left( \rho _{f}\rho _{i}\right) ^{\nu +%
\frac{1}{2}}\left( Q_{f}Q_{i}\right) ^{\nu }}\int \mathcal{D}Q\exp \left\{ 
\frac{i}{\hslash }\int \left[ \frac{\dot{Q}^{2}}{2}\right. \right.   \notag
\\
&&\left. \left. -\left[ \frac{\Omega ^{2}}{2}Q^{2}+\frac{\hslash ^{2}\left(
\beta ^{2}-\frac{1}{4}\right) }{2Q^{2}}+\bar{m}\bar{\rho}^{2}V\left( \bar{%
\rho}\,Q,\int\limits^{s}\bar{m}\bar{\rho}^{2}d\sigma \right) \right] \right]
ds\right\} .  \label{propagator-}
\end{eqnarray}

\section{Harmonic oscillator}

\label{sec3}

The result presented in Eq. (\ref{propagatorfinal}) is significant and
notably interesting due to its relative simplicity. In this section, we
demonstrate how this result can be applied, highlighting a noteworthy
application of substantial epistemological relevance: the harmonic
oscillator with time-dependent mass and frequency, 
\begin{equation}
V\left( x,t\right) =\frac{1}{2}m\left( t\right) \omega ^{2}\left( t\right)
x^{2}.
\end{equation}%
Employing the potential in Eq. (\ref{propagatorfinal}), the propagator takes
the form: 
\begin{eqnarray}
K &=&\frac{1}{\left( \rho _{f}\rho _{i}\right) ^{\nu +\frac{1}{2}}\left(
Q_{f}Q_{i}\right) ^{\nu }}\exp \left[ \frac{i}{2\hslash }\left( \frac{%
\overset{\cdot }{\bar{\rho}_{f}}}{\rho _{f}}Q_{f}^{2}-\frac{\overset{\cdot }{%
\bar{\rho}_{i}}}{\rho _{i}}Q_{i}^{2}\right) \right]   \notag \\
&&\times \left[ \int \mathcal{D}Q\exp \left\{ \frac{i}{\hslash }%
\int\limits_{s_{i}}^{s_{f}}\left[ \frac{\dot{Q}^{2}}{2}-\frac{\left( \Omega
^{2}+\bar{m}^{2}\bar{\rho}^{4}\omega ^{2}\right) Q^{2}}{2}-\frac{\hslash
^{2}\left( \alpha ^{2}-\frac{1}{4}\right) }{2Q^{2}}\right] ds\right\}
\right.   \notag \\
&&+\left. sng\left( x_{f}x_{i}\right) \int \mathcal{D}Q\exp \left\{ \frac{i}{%
\hslash }\int\limits_{s_{i}}^{s_{f}}\left[ \frac{\dot{Q}^{2}}{2}-\frac{%
\left( \Omega ^{2}+\bar{m}^{2}\bar{\rho}^{4}\omega ^{2}\right) Q^{2}}{2}-%
\frac{\hslash ^{2}\left( \beta ^{2}-\frac{1}{4}\right) }{2Q^{2}}\right]
ds\right\} \right] .  \label{29}
\end{eqnarray}%
We now impose a constraint on $\bar{\rho}$ by setting the global
time-dependent frequency in Eq. (\ref{29}) to a constant, 
\begin{equation}
\Omega ^{2}+\bar{m}^{2}\bar{\rho}^{4}\omega ^{2}=\text{constant}\equiv 1.
\label{omega}
\end{equation}%
Consequently, Eq.(\ref{newH}) becomes 
\begin{equation}
\ddot{\rho}+\frac{\dot{m}}{m}\dot{\rho}+\omega ^{2}\rho =\frac{1}{m^{2}\rho
^{3}},  \label{auxilary}
\end{equation}%
which is referred to as the auxiliary equation. By substituting Eq. (\ref%
{omega}) into Eq. (\ref{29}), we obtain the final expression for the
propagator of the time-dependent Dunkl-harmonic oscillator as: 
\begin{eqnarray}
K &=&\frac{1}{\left( \rho _{f}\rho _{i}\right) ^{\nu +\frac{1}{2}}\left(
Q_{f}Q_{i}\right) ^{\nu -\frac{1}{2}}}\exp \left[ \frac{i}{2\hslash }\left( 
\frac{\overset{\cdot }{\bar{\rho}_{f}}}{\rho _{f}}Q_{f}^{2}-\frac{\overset{%
\cdot }{\bar{\rho}_{i}}}{\rho _{i}}Q_{i}^{2}\right) \right]   \notag \\
&&\times \frac{1}{2i\hbar \sin \left( s_{f}-s_{i}\right) }\exp \left[ -\frac{%
\left( Q_{i}^{2}+Q_{f}^{2}\right) }{2i\hbar }\cot \left( s_{f}-s_{i}\right) %
\right]   \notag \\
&&\times \left( I_{\nu -\frac{1}{2}}\left( \frac{Q_{i}Q_{f}}{i\hbar \sin
\left( s_{f}-s_{i}\right) }\right) +sng\left( x_{f}x_{i}\right) I_{\nu +%
\frac{1}{2}}\left( \frac{Q_{i}Q_{f}}{i\hbar \sin \left( s_{f}-s_{i}\right) }%
\right) \right) ,  \label{20}
\end{eqnarray}%
where $I_{n}(x)$ is the modified Bessel function. Here, we have to underscore that the Dunkl approach extends classical propagators for harmonic oscillators by incorporating reflection symmetries, leading to solutions involving modified Bessel functions, instead of the standard trigonometric or exponential functions for the standard case, that provide parity-specific insights into the system's dynamics.

Returning to the original
variable $x$ and applying the decomposition outlined in Eq. (\ref{decom}),
we can express the even and odd kernel propagators as follows: 
\begin{eqnarray}
K_{+}\left( x_{f};t_{f};x_{i};t_{i}\right)  &=&\frac{1}{\left( \rho _{f}\rho
_{i}\right) \left\vert x_{f}x_{i}\right\vert ^{\nu -\frac{1}{2}}}\frac{\exp %
\left[ \frac{i}{2\hslash }\left( m_{f}\frac{\overset{\cdot }{\rho _{f}}}{%
\rho _{f}}x_{f}^{2}-m_{i}\frac{\overset{\cdot }{\rho _{i}}}{\rho _{i}}%
x_{i}^{2}\right) \right] }{2i\hbar \sin \left( S\right) }  \notag \\
&&\times \exp \left[ -\frac{\cot \left( S\right) }{2i\hbar }\left( \frac{%
x_{i}^{2}}{\rho _{i}^{2}}+\frac{x_{f}^{2}}{\rho _{f}^{2}}\right) \right]
I_{\nu -\frac{1}{2}}\left( \frac{\left\vert x_{f}x_{i}\right\vert }{i\hbar
\rho _{i}\rho _{f}\sin \left( S\right) }\right) ,
\end{eqnarray}%
\begin{eqnarray}
K_{-}\left( x_{f};t_{f};x_{i};t_{i}\right)  &=&\frac{1}{\left( \rho _{f}\rho
_{i}\right) \left\vert x_{f}x_{i}\right\vert ^{\nu -\frac{1}{2}}}\frac{\exp %
\left[ \frac{i}{2\hslash }\left( m_{f}\frac{\overset{\cdot }{\rho _{f}}}{%
\rho _{f}}x_{f}^{2}-m_{i}\frac{\overset{\cdot }{\rho _{i}}}{\rho _{i}}%
x_{i}^{2}\right) \right] }{2i\hbar \sin \left( S\right) }  \notag \\
&&\times \exp \left[ -\frac{\cot \left( S\right) }{2i\hbar }\left( \frac{%
x_{i}^{2}}{\rho _{i}^{2}}+\frac{x_{f}^{2}}{\rho _{f}^{2}}\right) \right]
I_{\nu +\frac{1}{2}}\left( \frac{\left\vert x_{f}x_{i}\right\vert }{i\hbar
\rho _{i}\rho _{f}\sin \left( S\right) }\right) ,
\end{eqnarray}%
where 
\begin{equation}
S=\int\limits_{t_{i}}^{t_{f}}\frac{d\sigma }{\rho ^{2}\left( \sigma \right)
m\left( \sigma \right) }.
\end{equation}%
Setting 
\begin{equation}
z=e^{-iS},\qquad X=\frac{1}{\sqrt{\hbar }}\frac{x_{i}}{\rho _{i}},\qquad Y=%
\frac{1}{\sqrt{\hbar }}\frac{x_{f}}{\rho _{f}},
\end{equation}%
in the expression for the kernel propagators given in Eq. (\ref{20}), and
employing the Hille-Hardy formula \cite{Rosenblum}, 
\begin{equation}
\underset{n=0}{\overset{\infty }{\sum }}\frac{n!L_{n}^{\theta }\left(
X\right) L_{n}^{\theta }\left( Y\right) }{\Gamma \left( n+\theta +1\right) }%
Z^{n}\exp \left[ -\frac{X+Y}{2}\right] =\frac{\left( XYZ\right) ^{-\frac{%
\theta }{2}}}{1-Z}\exp \left[ -\frac{\left( X+Y\right) \left( 1+Z\right) }{%
1-Z}\right] I_{\theta }\left( \frac{2\sqrt{XYZ}}{1-z}\right) ,
\label{HJ functions}
\end{equation}%
we derive the even and odd spectral decompositions of kernel propagators: 
\begin{equation}
K(x_{f};t_{f},x_{i};t_{i})=\underset{n}{\sum }\left\{ \Psi _{n,+}\left(
x_{f},t_{f}\right) \Psi _{n,+}^{^{\ast }}\left( x_{i},t_{i}\right) +\Psi
_{n,-}\left( x_{f},t_{f}\right) \Psi _{n,-}^{^{\ast }}\left(
x_{i},t_{i}\right) \right\} ,
\end{equation}%
where  
\begin{eqnarray}
\Psi _{n,+}\left( x,t\right)  &=&\sqrt{\frac{n!}{\hslash ^{\nu +\frac{1}{2}%
}\Gamma \left( n+\nu +\frac{1}{2}\right) \rho ^{2\nu +1}}}L_{n}^{\nu -\frac{1%
}{2}}\left( \frac{x^{2}}{\hbar \rho ^{2}}\right)   \notag \\
&&\times \exp \left[ \frac{i}{2\hslash }\left( m\rho \dot{\rho}-1\right) 
\frac{x^{2}}{\rho ^{2}}-i\left( 2n+\nu +\frac{1}{2}\right) \int\limits^{t}%
\frac{d\sigma }{\rho ^{2}\left( \sigma \right) m\left( \sigma \right) }%
\right] ,
\end{eqnarray}%
is the even wave functions and  
\begin{eqnarray}
\Psi _{n,-}\left( x,t\right)  &=&\sqrt{\frac{n!}{\hbar ^{\nu +\frac{3}{2}%
}\Gamma \left( n+\nu +\frac{3}{2}\right) \rho ^{^{2\nu +3}}}}xL_{n}^{\nu +%
\frac{1}{2}}\left( \frac{x^{2}}{\hbar \rho ^{2}}\right)   \notag \\
&&\times \exp \left[ \frac{i}{2\hslash }\left( m\rho \dot{\rho}-1\right) 
\frac{x^{2}}{\rho ^{2}}-i\left( 2n+\nu +\frac{3}{2}\right) \int\limits^{t}%
\frac{d\sigma }{\rho ^{2}\left( \sigma \right) m\left( \sigma \right) }%
\right] ,
\end{eqnarray}%
the odd wave functions,  $L_{n}^{\nu \mp \frac{1}{2}}$ are the associated
Legendre or Laguerre polynomial. It is noteworthy that the eigenfunctions we
obtained coincide with those derived in \cite{Benchikha2} through the
Lewis-Riesenfeld method. Then, with the help of the deformed exponential
function $E_{\nu }$ \cite{Rosenblum}, defined as: 
\begin{equation}
E_{\nu }\left( z\right) =\Gamma \left( \nu +\frac{1}{2}\right) \left( \frac{2%
}{\left\vert z\right\vert }\right) ^{\nu -\frac{1}{2}}\left( I_{\nu -\frac{1%
}{2}}\left( \left\vert z\right\vert \right) +sng\left( z\right) I_{\nu +%
\frac{1}{2}}\left( \left\vert z\right\vert \right) \right) ,
\end{equation}%
we express the propagator of the time-dependent Dunkl-harmonic oscillator as
follows: 
\begin{eqnarray}
K &=&\frac{1}{\Gamma \left( \nu +\frac{1}{2}\right) }\left( \frac{1}{2i\hbar
\rho _{i}\rho _{f}\sin \left( S\right) }\right) ^{\nu +\frac{1}{2}}\exp %
\left[ -\frac{\cot \left( S\right) }{2i\hbar }\left( \frac{x_{i}^{2}}{\rho
_{i}^{2}}+\frac{x_{f}^{2}}{\rho _{f}^{2}}\right) \right]   \notag \\
&&\times \times \exp \left[ \frac{i}{2\hslash }\left( m_{f}\frac{\overset{%
\cdot }{\rho _{f}}}{\rho _{f}}x_{f}^{2}-m_{i}\frac{\overset{\cdot }{\rho _{i}%
}}{\rho _{i}}x_{i}^{2}\right) \right] E_{\nu }\left( \frac{x_{i}x_{f}}{%
i\hbar \rho _{i}\rho _{f}\sin \left( S\right) }\right) ,
\label{HOpropagator}
\end{eqnarray}

where $S=\int\limits_{t_{i}}^{t_{f}}\frac{d\sigma }{\rho ^{2}\left( \sigma
\right) m\left( \sigma \right) }.$ It is important to note that for the case
where the mass and frequency are constants, i.e. $m\left( t\right) =m$ and \ 
$\omega \left( t\right) =1$, we recover the propagator of the
one-dimensional harmonic oscillator within the framework of Wigner-Dunkl
quantum mechanics \cite{Junker},%
\begin{eqnarray}
K &=&\frac{\Gamma \left( \nu +1\right) }{\sqrt{2\pi }\Gamma \left( 2\nu
+1\right) }\left( \frac{1}{i\hbar \sin \left( t_{f}-t_{i}\right) }\right)
^{\nu +\frac{1}{2}}E_{\nu }\left( \frac{mx_{i}x_{f}}{i\hbar \sin \left(
t_{f}-t_{i}\right) }\right)   \notag \\
&&\times \exp \left[ \frac{im}{2\hslash }\left( x_{f}^{2}-x_{i}^{2}\right) -%
\frac{\cot \left( t_{f}-t_{i}\right) }{2i\hbar }\left(
x_{i}^{2}+x_{f}^{2}\right) \right] .
\end{eqnarray}

\section{Applications}

\label{sec4} At this stage, it is necessary to examine specific models of
time-varying harmonic oscillators to validate the analytical result
presented in Eq. (\ref{20}). These models should provide well-defined
expressions for $m\left( t\right) $ and $\omega \left( t\right)$ that yield
exact solutions. In the following subsection, we explore two examples: the
Caldirola-Kanai oscillator and a harmonic oscillator with a strongly
pulsating mass.

\subsection{Caldirola-Kanai Oscillator}

Let us begin by examining a harmonic oscillator with a constant frequency
and a mass that varies with time in the form of 
\begin{equation*}
m\left( t\right) =m_{0}e^{kt},
\end{equation*}%
where $m_{0}$ is a real number, and $k$ is the constant damping coefficient.
In this case, the Dunkl-Caldirola-Kanai Hamiltonian for this system can be
expressed as: 
\begin{equation}
H=-\frac{\hbar ^{2}}{2m_{0}}\left( \frac{\partial ^{2}}{\partial x^{2}}+%
\frac{2\nu }{x}\frac{\partial }{\partial x}-\frac{\nu \left( 1-R\right) }{%
x^{2}}\right) e^{-kt}+\frac{1}{2}m_{0}\omega _{0}^{2}e^{kt}x^{2},  \label{30}
\end{equation}%
and the solution of the auxiliary equation, given in Eq. (\ref{auxilary}),
takes the form 
\begin{equation}
\rho =\frac{e^{-\frac{k}{2}t}}{\sqrt{m_{0}\mu }},  \label{31}
\end{equation}%
where $\mu $ is the reduce frequency 
\begin{equation}
\mu =\sqrt{\omega _{0}^{2}-\frac{k^{2}}{4}}.  \label{33}
\end{equation}%
By employing a time transformation defined in Eq. \eqref{time transf}, we
obtain an exact expression for the propagator $K\left(
x_{f};t_{f},x_{i};t_{i}\right) $ as follows: 
\begin{eqnarray}
K &=&\frac{1}{\Gamma \left( \nu +\frac{1}{2}\right) }\left( \frac{m_{0}\mu
e^{\frac{k}{2}\left( t_{f}+t_{i}\right) }}{2i\hbar \sin \mu \left(
t_{f}-t_{i}\right) }\right) ^{\nu +\frac{1}{2}}E_{\nu }\left( \frac{m_{0}\mu
e^{\frac{k}{2}\left( t_{f}+t_{i}\right) }x_{i}x_{f}}{i\hbar \sin \mu \left(
t_{f}-t_{i}\right) }\right)   \notag \\
&&\times \exp \left[ \frac{im_{0}\mu }{2\hbar }\left(
e^{kt_{f}}x_{f}^{2}+e^{kt_{i}}x_{i}^{2}\right) \cot \mu \left(
t_{f}-t_{i}\right) \frac{-im_{0}k}{4\hbar }\left(
e^{kt_{f}}x_{f}^{2}-e^{kt_{i}}x_{i}^{2}\right) \right] .
\end{eqnarray}%
By inserting this propagator in Eq. (\ref{HJ functions}), we readily derive
the wave functions. The even and odd wave functions are expressed,
respectively, in terms of Legendre polynomials as follows: 
\begin{eqnarray}
\psi _{n,\nu }^{+}\left( x,t\right)  &=&\sqrt{\tfrac{n!\left( \frac{m_{0}\mu 
}{\hslash }\right) ^{\nu +\frac{1}{2}}}{\Gamma \left( n+\nu +\frac{1}{2}%
\right) }}L_{n}^{\nu -\frac{1}{2}}\left( \frac{\mu m_{0}e^{kt}}{\hbar }%
x^{2}\right)   \notag \\
&&\times \exp \left[ -\frac{m_{0}e^{kt}}{2\hslash }\left( \frac{ik}{2}+\mu
\right) x^{2}+\left[ \frac{k}{2}\left( \nu +\frac{1}{2}\right) -i\mu \left(
2n+\nu +\frac{1}{2}\right) \right] t\right] ,
\end{eqnarray}%
\begin{eqnarray}
\psi _{n,\nu }^{-}\left( x,t\right)  &=&\sqrt{\tfrac{n!\left( \frac{m_{0}\mu 
}{\hslash }\right) ^{\nu +\frac{3}{2}}}{\Gamma \left( n+\nu +\frac{3}{2}%
\right) }x}L_{n}^{\nu +\frac{1}{2}}\left( \frac{\mu m_{0}e^{kt}}{\hbar }%
x^{2}\right)   \notag \\
&&\times \exp \left[ -\frac{m_{0}e^{kt}}{2\hslash }\left( \frac{ik}{2}+\mu
\right) x^{2}+\left[ \frac{k}{2}\left( \nu +\frac{3}{2}\right) -i\mu \left(
2n+\nu +\frac{3}{2}\right) \right] t\right] .
\end{eqnarray}%
We note that this result is consistent with that obtained in \cite%
{Benchikha2}.

\subsection{The harmonic oscillator with strongly pulsating mass}

\bigskip The harmonic oscillator with a strongly pulsating mass can be used
to model the interaction between the electromagnetic field in a Fabry-Perot
cavity and a reservoir of resonant two-level atoms. The periodic emission
and reabsorption of photons can be described by an oscillator with energy
that fluctuates periodically. This system can be represented by an
oscillator with a periodically fluctuating energy, which corresponds to a
mass that varies periodically over time as,%
\begin{equation}
m\left( t\right) =m_{0}\cos ^{2}\upsilon t,  \label{mapu}
\end{equation}%
where $m_{0}$ is the constant, $\upsilon $ represents the frequency of mass.
In this case, the Hamiltonian transforms into 
\begin{equation}
H=-\frac{\hbar ^{2}}{2m_{0}}\left( \frac{\partial ^{2}}{\partial x^{2}}+%
\frac{2\nu }{x}\frac{\partial }{\partial x}-\frac{\nu \left( 1-s\right) }{%
x^{2}}\right) \sec ^{2}\upsilon t+\frac{1}{2}m_{0}\cos ^{2}\upsilon t\omega
_{0}^{2}x^{2}.  \label{nh}
\end{equation}%
By inserting the time-dependent mass into the auxiliary Eq.(\ref{auxilary}),
we obtain 
\begin{equation}
\rho =\frac{1}{\sqrt{m_{0}\eta }\cos \upsilon t},
\end{equation}%
where $\eta =\sqrt{\omega _{0}^{2}+\upsilon ^{2}}$, is the augmented
frequency. By inserting Eqs. (\ref{mapu}) and (\ref{nh}) into the general
propagator, we can derive the time-dependent propagator for a Dunkl-harmonic
oscillator with a strongly pulsating mass as follows,

\begin{eqnarray}
K\left( x_{f};t_{f},x_{i};t_{i}\right)  &=&\frac{1}{\Gamma \left( \nu +\frac{%
1}{2}\right) }\left( \frac{m_{0}\eta \cos \upsilon t_{i}\cos \upsilon t_{f}}{%
2i\hbar \sin \eta \left( t_{f}-t_{i}\right) }\right) ^{\nu +\frac{1}{2}%
}E_{\nu }\left( \frac{m_{0}\eta x_{i}x_{f}\cos \upsilon t_{i}\cos \upsilon
t_{f}}{i\hbar \sin \eta \left( t_{f}-t_{i}\right) }\right)   \notag \\
&&\times \exp \left\{ \frac{im_{0}\eta }{2\hbar }\left( x_{f}^{2}\cos
^{2}\upsilon t_{f}+x_{i}^{2}\cos ^{2}\upsilon t_{i}\right) \cot \eta \left(
t_{f}-t_{i}\right) \right\}   \notag \\
&&\times \exp \left\{ \frac{im_{0}\upsilon }{2\hbar }\left( x_{f}^{2}\cos
^{2}\upsilon t_{f}\tan \upsilon t_{f}-x_{i}^{2}\cos ^{2}\upsilon t_{i}\tan
\upsilon t_{i}\right) \right\} .
\end{eqnarray}%
This propagator can be simplified by setting $\upsilon =0$, reducing the
result to the propagator of the stationary Dunkl harmonic oscillator \cite%
{Benchikha2}.

The results for the Dunkl-Caldirola-Kanai oscillator reveal that the exponentially varying mass effectively models quantum damping effects, highlighting the system's relevance to dissipative quantum mechanics, while the harmonic oscillator with a strongly pulsating mass captures periodic energy fluctuations, providing insights into systems with time-dependent interactions such as cavity quantum electrodynamics.


\subsection{Comparison of Dunkl Formalism and Traditional Approaches}

Before concluding the manuscript, it is important to highlight the key distinctions between the Dunkl formalism and traditional methods for propagators. These differences are summarized in the following table:

\begin{table}[h!]
\centering
\caption{Comparison of Dunkl Formalism and Traditional Approaches for Propagators}
\begin{tabular}{|p{5cm}|p{5cm}|p{5cm}|}
\hline
\textbf{Aspect}                & \textbf{Traditional Approach}                    & \textbf{Dunkl Formalism}                        \\ \hline
Symmetry Consideration         & Parity symmetry not explicitly integrated        & Incorporates reflection symmetry explicitly     \\ \hline
Mathematical Tools             & Standard derivatives and trigonometric/exponential functions & Dunkl operator and modified Bessel, Legendre and Laguerre functions   \\ \hline
Propagator Decomposition       & Not applicable                                  & Even and odd parity components                 \\ \hline
Applications                   & General quantum systems                         & Systems with parity/reflection symmetries       \\ \hline
\end{tabular}
\label{tab:dunkl_comparison}
\end{table}

\section{Conclusion}\label{concl}

Time-dependent quantum mechanics has become a pivotal area of research due to its applications across diverse fields, including plasma physics, quantum optics, and quantum chemistry. While the path integral method is widely used in time-dependent quantum systems, its application within the Dunkl formalism remains largely unexplored. In this study, we extended the path integral approach to time-dependent Dunkl quantum mechanics by reducing it to a stationary form through explicit generalized canonical and time transformations.

More precisely, we derived a general expression for the propagator and developed explicit formulations for the Dunkl harmonic oscillator. A particular focus was placed on two analytically solvable cases involving harmonic oscillators with time-dependent mass: the Dunkl-Caldirola-Kanai oscillator and a harmonic oscillator with a strongly pulsating mass. 

The Dunkl-Caldirola-Kanai oscillator, characterized by an exponentially varying mass, effectively models quantum damping effects. Moreover, the harmonic oscillator with a strongly pulsating mass captures periodic energy fluctuations, relevant for systems such as cavity quantum electrodynamics. These results underscore the unique ability of the Dunkl formalism to incorporate reflection symmetries and handle time-dependent parameters, leading to parity-specific propagators and wave functions expressed in terms of modified Bessel functions, instead of the standard trigonometric or exponential functions for the standard case.

Beyond theoretical insights, the practical implications of this work are significant. The Dunkl formalism provides a framework for studying quantum dissipative systems, where exponentially varying mass can model damping effects. Furthermore, it is well-suited for investigating non-Hermitian quantum mechanics, where reflection symmetries could reveal new physical phenomena. The exact solutions derived here could also serve as benchmarks for numerical simulations of time-dependent quantum systems, aiding in the development of more efficient computational techniques.

Future studies could explore the interplay between Dunkl mechanics and non-Hermitian quantum systems, potentially uncovering new dynamical regimes. Additionally, extending the formalism to multi-dimensional systems or PT-symmetric systems would broaden its applicability and relevance.

\section*{Acknowledgments}

The authors extend their sincere gratitude to the anonymous reviewer for his/her insightful comments and constructive feedback, which have significantly contributed to the improvement of this manuscript. B. C. L. is grateful to Excellence Project PřF UHK 2211/2023-2024 for the financial support.

\section*{Data Availability Statement} 

Data sharing is not applicable to this article as no new data were created or analyzed in this study.

\section*{Conflict of Interests} 

The authors have no conflicts to disclose.

\end{document}